\documentclass[universe,article,accept,pdftex,moreauthors]{Definitions/mdpi}

\usepackage{geometry}
\RequirePackage[T1]{fontenc}
\usepackage[utf8]{inputenc}
\usepackage[english]{babel}
\usepackage{graphicx}
\usepackage{bm}
\usepackage{url}
\usepackage{physics}
\usepackage{amssymb}
\usepackage{float}
\usepackage{booktabs}
\usepackage{caption}
\usepackage{subcaption}
\microtypesetup{nopatch=footnote}

\firstpage{1} 
\pubvolume{1}
\issuenum{1}
\articlenumber{0}
\pubyear{2025}
\copyrightyear{2025}
\externaleditor{} % More than 1 editor, please add `` and '' before the last editor name
\datereceived{2 January 2025} 
\daterevised{3 February 2025} 
\dateaccepted{10 February 2025} 
\datepublished{} 
\hreflink{https://\\doi.org/}

\newcommand{\ham}[0]{\mathcal{H}}

\Title{Fate of Mixmaster Chaos in a Deformed Algebra Framework}
\TitleCitation{Fate of Mixmaster Chaos in a Deformed Algebra Framework}

\Author{Gabriele %MDPI: Please carefully check the accuracy of names and affiliations.
 Barca \textsuperscript{1,}*\orcidA{} and Eleonora Giovannetti \textsuperscript{2}\orcidB{}}
\AuthorNames{Gabriele Barca and Eleonora Giovannetti}
\AuthorCitation{Barca, G.; Giovannetti, E.}

\address{\textsuperscript{1} \quad School of Mathematical and Physical Sciences, University of Sheffield, Hicks Building, Hounsfield Road, Sheffield S3 7RH, UK\\
\textsuperscript{2} \quad Department of Physics, La Sapienza University of Rome, P.le Aldo Moro, 00185 Roma, Italy; eleonora.giovannetti@uniroma1.it}
\corres{Correspondence: gabriele\_barca001@ehu.eus} %MDPI: This email address is different from the submitting system. Please confirm which one is correct. %Answer (G. Barca): this is the correct institutional email address. It is different than the submitting one because I was in the middle of changing institutions: I knew that my previous one was going to be deleted (and indeed it already has), but the new one had not been assigned yet, so I used my personal address to submit the manuscript, but now that I have an institutional one it is better to use that. I will still keep the old affiliation because all of the work (and also the first submission) was completed there.

\abstract{We analyze the anisotropic Bianchi models, and in particular the Bianchi Type IX known as the Mixmaster universe, where the Misner anisotropic variables obey Deformed Commutation Relations inspired by Quantum Gravity theories. We consider three different deformations, two of which have been able to remove the initial singularity similarly to Loop Quantum Cosmology when implemented in the single-volume variable. %EE: please check that the intended meaning has been retained
	Here, the two-dimensional algebras naturally implement a form of Non-Commutativity between the space variables that affects the dynamics of the anisotropies. In particular, we implement the modifications in their classical limit, where the Deformed Commutators become Deformed Poisson Brackets. We derive the modified Belinskii--Khalatnikov--Lifshitz map in all the three cases, and we study the fate of the chaotic behavior that the model classically presents. Depending on the sign of the deformation, the dynamics will either settle into oscillations between two almost-constant angles, or stop reflecting after a finite number of iterations and reach the singularity as one last simple Kasner solution. In either case, chaos is removed.}
\keyword{Mixmaster universe; alternative quantum mechanics; non-commutativity; chaos}

\begin{document}
\section{Introduction}
One of the most general predictions of Relativistic Cosmology is the existence of singularities~\cite{Singularity1,Singularity2}, where curvature invariants diverge and General Relativity ceases to be predictive. However, it is expected that in the high-energy regimes close to the singularities, quantum effects must come into~play.

The two most successful proposals for formulating Quantum Gravity (QG) theories are Loop Quantum Gravity (LQG)~\cite{GeneralLQG1,GeneralLQG2,GeneralLQG3} {(see~\cite{LQGreview2,LQGreview1} for recent reviews)} and String Theories (STs)~\cite{STbook,ST1,ST2}. However, their implementation in the cosmological setting, resulting  in the frameworks of Loop Quantum Cosmology (LQC)~\cite{BojowaldOriginalLQC,LQC2011Review} and Brane Cosmology (BC)~\cite{BraneWorldGravity,BraneCosmology1,BraneCosmology2}, respectively, is rather complicated, especially for LQC~\cite{LQCproblems1,LQCproblems2}. As~a consequence, the~new formalism of Deformed Commutation Relations (DCRs) has been developed as a way to easily implement quantum gravitational effects to any Hamiltonian~system.

The first and most famous DCR is the Generalized Uncertainty Principle (GUP) representation by Kempf, Mangano and Mann (KMM-GUP), which is inspired by String Theories. By~adding a higher-order momentum term in the standard commutator, the~corresponding uncertainty relations are modified in such a way that an absolute minimal uncertainty on position appears~\cite{KMM,Maggiore93,ScardigliGUP}. This is a reminder that it is not possible to probe below the string scale, and~the GUP representation is a simple way to introduce the concept of a minimal length expected in QG~theories.

The DCRs can then be generalized to other functions of the momentum; different generalizations are possible, but~it is these that are usually able to introduce the QG cut-offs of interest {\cite{Hossenfelder}}. We use three deformations in particular~\cite{IJGMMP}: the Loop Algebra, which introduces a momentum (i.e. energy) cut-off and is able to reproduce effective LQG by replacing singularities with a Bounce; the Brane Algebra, an~extended version of the KMM representation, which reproduces Brane Cosmology; and a hybrid between the Loop and the KMM Algebras which we will call the Loop Uncertainty Principle (LUP), which removes singularities through asymptotes. In~higher space dimensions, there are some problems, and~therefore, for the time being, we will restrict ourselves to the classical limit of these representations, where the DCRs become Deformed Poisson Brackets. However, due to the Jacobi identities, a~Non-Commutativity between the position variables will emerge naturally~\cite{Maggiore2021,SebyGUPBianchi}. Therefore, these Deformed Algebras represent a straightforward way of introducing QG effects and Non-Commutativity at an effective level in a simple and natural~way.

In this paper, however, we are not interested in the removal of cosmological singularities. Rather, we are interested in studying how the chaotic dynamics of the anisotropic Bianchi IX universe~\cite{Misner2} is affected by these QG corrections, in~a similar vein to what has recently been achieved for the original KMM formulation~\cite{SebyMixmaster}. We will use Misner variables $(\alpha,\beta_+,\beta_-)$, and~it is well known that it is not possible to remove singularities when using the logarithmic variable $\alpha$ instead of scale factors or other powers; indeed, the nature of the resulting modified dynamics seems to strongly depend on what variable is chosen to be modified~\cite{Mantero,GiovannettiMixmaster,Mandini,Review}.

The Bianchi models are homogeneous but anisotropic cosmological models. There are nine classes, each corresponding to the Lie Algebra of a different three-dimensional group. The~interest of studying them lies in their generality, in~particular the Bianchi type VIII and Type IX models, which present  chaotic dynamics towards the singularity. Indeed, Bianchi IX, also dubbed Mixmaster~\cite{Misner2}, is the most general homogeneous cosmological model that presents an isotropic limit. In~Misner variables, the~Mixmaster dynamics are those of a point particle in the $(\beta_+,\beta_-)$ plane, constrained in an exponentially steep potential well with the symmetry of an equilateral triangle; the particle universe will keep reflecting off these potential walls indefinitely towards the singularity, and~the dynamics acquire a chaotic character. Besides~being the starting point for the general inhomogeneous cosmological solution~\cite{BKL,PC}, its chaotic dynamics make it the perfect arena to test and study quantum corrections and quantum gravitational effects. {Indeed, there are many different studies of the Bianchi IX model (and of its inhomogeneous extension) with different approaches; for some examples related to LQC, see~\cite{CrinoPintaudi,MixmasterReview,GiovannettiMixmaster,LQCBIX,LQCWilsonEwing,MontaniInhomogeneousMixmaster,Antonini}.}

Therefore, here, we will consider the Mixmaster model in terms of Misner variables. After~performing a reduction that allows us to use the isotropic variable $\alpha$ as time~\cite{ADM}, we will deform the anisotropy variables $\beta_\pm$, which together form a 2D space that will therefore present Non-Commutativity between the two variables $\beta_+$ and $\beta_-$. As~a consequence, the~reflection law against the potential walls will be modified, and~the chaotic dynamics will change. In~particular, we find that the sign of the quantum gravitational correction in the Deformed Poisson Brackets will determine the type of change. When there is a plus sign, the~trajectory will oscillate between two almost-constant angles, similarly to~\cite{SebyMixmaster}; on the other hand, when there is a minus sign, the~number of reflections is finite and the particle universe slows down, until~no reflections happen anymore and the singularity is reached with Kasner-like free-particle dynamics~\cite{Kasner}. Furthermore, similar initial conditions will yield similar dynamics. Therefore, in~all cases, we can claim that QG corrections are able to remove chaos. This study represents an advancement towards the characterization of QG effects in a more general cosmological~setting.

The manuscript is organized as follows. In~Section~\ref{BIX}, we present the classical Bianchi IX model, with~particular focus on its chaotic dynamics. In~Section~\ref{Algebras}, we introduce the framework and main properties of the Deformed Commutation Relations. They will then be used in Section~\ref{DeformedBianchi} to study the fate of chaos and to derive the modified Belinskii--Khalatnikov--Lifshitz maps that classically describe the chaotic dynamics. We then conclude the manuscript with a summary and discussion in Section~\ref{Concl}. We will use natural units $\hslash=c=8\pi G=1$.
\section{The Bianchi IX~Model}
\label{BIX}
In this section, we present the most important results about the Bianchi IX dynamics towards the cosmological singularity, i.e.,~the Mixmaster model~\cite{Misner1,Misner2}. It is the most generic homogeneous but anisotropic model and its relevance lies in the possibility of using this description to construct the generic cosmological solution~\cite{BKL,PC}.

The line element in the Misner variables is
\begin{equation}
ds^2=-N(t)^2dt^2+\frac{1}{4}e^{2\alpha}(e^{2\beta})_{ij}\sigma_i\sigma_j\,,
\end{equation}
where $N(t)$ is the Lapse function parametrizing the freedom of choosing different time variables, $\sigma_i$ has three differential forms depending on the Euler angles of the $SO(3)$ group of symmetry, %EE: please check that the intended meaning has been retained
and $\beta_{ij}$ is a diagonal, traceless matrix that can be parametrized in terms of the two independent variables $(\beta_+,\beta_-)$ as ${\beta_{ij}=\text{diag}(\beta_++\sqrt{3}\beta_-,\beta_+-\sqrt{3}\beta_-,-2\beta_+)}$. The~Misner variables $(\alpha,\beta_\pm)$ represent a very convenient choice to parametrize the Bianchi IX line element, since the information on the evolution of the universe volume $v=e^{3\alpha}$ is separated from the anisotropic content of the model, parametrized by the variables $(\beta_+,\beta_-)$. Moreover, in~the Misner variables, the Hamiltonian of the model is diagonal in its kinetic part, as~one can see from the below formula:
\begin{equation}
\label{H}
\mathcal{H}=\frac{N}{3(8\pi)^2}e^{-3\alpha}\bigg(-p_{\alpha}^2+p_+^2+p_-^2+3(4\pi)^4e^{4\alpha}V(\beta_\pm)\bigg)=0\,,
\end{equation}
where $(p_\alpha,p_\pm)$ are the conjugate momenta to $(\alpha,\beta_\pm)$, respectively. Due to spatial curvature\endnote{The Bianchi IX model reduces to the closed FLRW universe in the isotropic limit.}, there is also a potential term $V(\beta_\pm)$ depending only on the anisotropy variables, whose explicit form is
%MDPI: We revised footnote to Notes.

\begin{equation}
\label{V}
V(\beta_\pm)=2e^{4\beta_+}\big(\cosh(4\sqrt{3}\beta_-)-1	\big)-4e^{2\beta_+}\cosh(2\sqrt{3}\beta_-)+e^{-8\beta_+}\,.
\end{equation}

\begin{figure}[H]
    %\centering
    \includegraphics[width=0.5\linewidth]{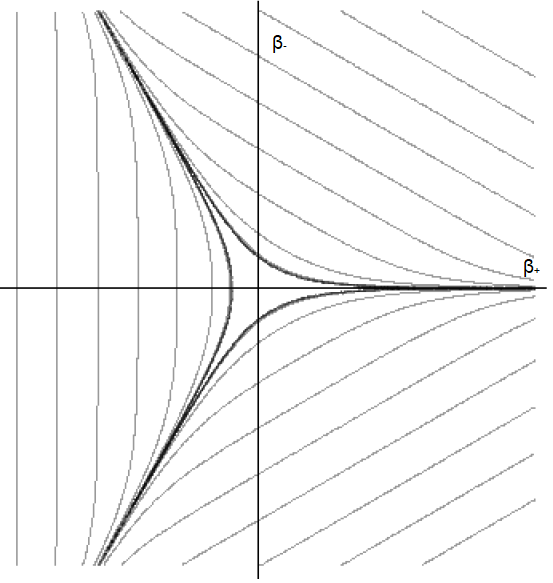}
    \caption{Equipotential %MDPI: Please change the hyphen (-) into minus sign ($-$, "U+2212"). 
 lines of the Bianchi IX potential in the $(\beta_+,\beta_-)$ plane. Credits:~\cite{GiovannettiMixmaster}.}
    \label{potential}
\end{figure}

This %MDPI: We added an indentation, please check. Answer: if possible, we think that all added indentations should be removed (except where specified), because those sentences should still be part of the same paragraph (and indeed most of them start with "this way" or refer to "this result"), and not start another paragraph.
 term defines a closed potential well with exponentially steep walls and the symmetry of an equilateral curvilinear triangle with open corners at infinity, as~we can see in Figure~\ref{potential}. In~addition, potential walls move outward while approaching the cosmological singularity ($\alpha\rightarrow-\infty$), due to the term $e^{4\alpha}$ in \eqref{H}.

The most interesting properties of the Bianchi IX dynamics emerge when applying the Arnowitt--Deser--Misner (ADM) reduction~\cite{ADM}. In~particular, if~we choose $\alpha$ as an internal time and hence solve the Hamiltonian constraint with respect to the variable $p_{\alpha}$, the~Mixmaster dynamics becomes isomorphic to that of a two-dimensional point particle, i.e.,~a point universe, that moves inside a potential well in which the spatial directions are identified by the two anisotropy variables $(\beta_+,\beta_-)$. The~reduced ADM Hamiltonian is
\begin{equation}
\label{HADM}
\mathcal{H}_{\text{ADM}}:=-p_{\alpha}=\sqrt{p_+^2+p_-^2+3(4\pi)^4e^{4\alpha}V(\beta_\pm)}\,
\end{equation}
and the classical dynamics of the model can be studied by solving the corresponding Hamilton equations\endnote{The dot denotes the derivative with respect to $\alpha$.}
\begin{equation}
\begin{cases}
	\label{EqHam}
\dot{\beta}_\pm=\displaystyle\frac{p_\pm}{\mathcal{H}_{\text{ADM}}}\,,
	\\
\dot{p}_\pm=\displaystyle-\frac{e^{4\alpha}}{2\mathcal{H}_{\text{ADM}}}3(4\pi)^4\frac{\partial V(\beta_\pm)}{\partial\beta_\pm}\,,
	\\
\dot{\mathcal{H}}_{\text{ADM}}=\displaystyle\frac{6(4\pi)^4e^{4\alpha}V(\beta_\pm)}{\mathcal{H}_{\text{ADM}}}\,.
\end{cases}
\end{equation}

\vspace{-9pt}

Because of the steepness of the walls, the~point universe acts as a free particle for most of the motion, except~when it is reflected off one of the three walls. In~particular, using the free-particle approximation\endnote{Due to the regression of the walls, the~free-particle approximation becomes more and more valid as the particle goes towards the singularity.}, i.e.,~reducing to the Bianchi I model
\begin{equation}
\mathcal{H}^{I}=\sqrt{p_+^2+p_-^2}\,,
\end{equation}
(in which $V(\beta_\pm)=0$), we can derive the velocity of the point universe from Hamilton equations, i.e.,~the anisotropy velocity
\begin{equation}
\label{beta}
\dot{\beta}\equiv\sqrt{{\dot{\beta}_+}^2+{\dot{\beta}_-}^2}=1\,.
\end{equation}

Moreover, %MDPI: We added an indentation, please check.
 studying the region in which the potential term is relevant gives information about the position of the walls and their velocity $|\dot{\beta}_{\text{wall}}|=\frac{1}{2}$. Furthermore, combining integrals of motion, it can be demonstrated that every reflection occurs according to the law
\begin{equation}
\label{reflection}
\frac{1}{2}\sin(\theta_i+\theta_f)=\sin\theta_i-\sin\theta_f\,,
\end{equation}
where $\theta_i$ and $\theta_f$ are the incidence and reflection angles computed with respect to the potential wall's normal. This reflection law is equivalent to the so-called Belinskii--Khalatnikov--Lifshitz (BKL) map for the Kasner indices that appears when using directional scale factors instead of Misner variables~\cite{PC,Kasner}. Here, we will use the acronym BKL to refer interchangeably to both the reflection law and the BKL~map.

By imposing the condition $|\dot{\beta}_{\perp}|>\dot{\beta}_{\text{wall}}$, one finds that the maximum incidence angle~is
\begin{equation}
\theta_{max}\equiv\arccos\bigg(\frac{1}{2}\bigg)=\frac{\pi}{3}\,.
\end{equation}

\begin{figure}[H]
  %  \centering
    \includegraphics[width=\linewidth]{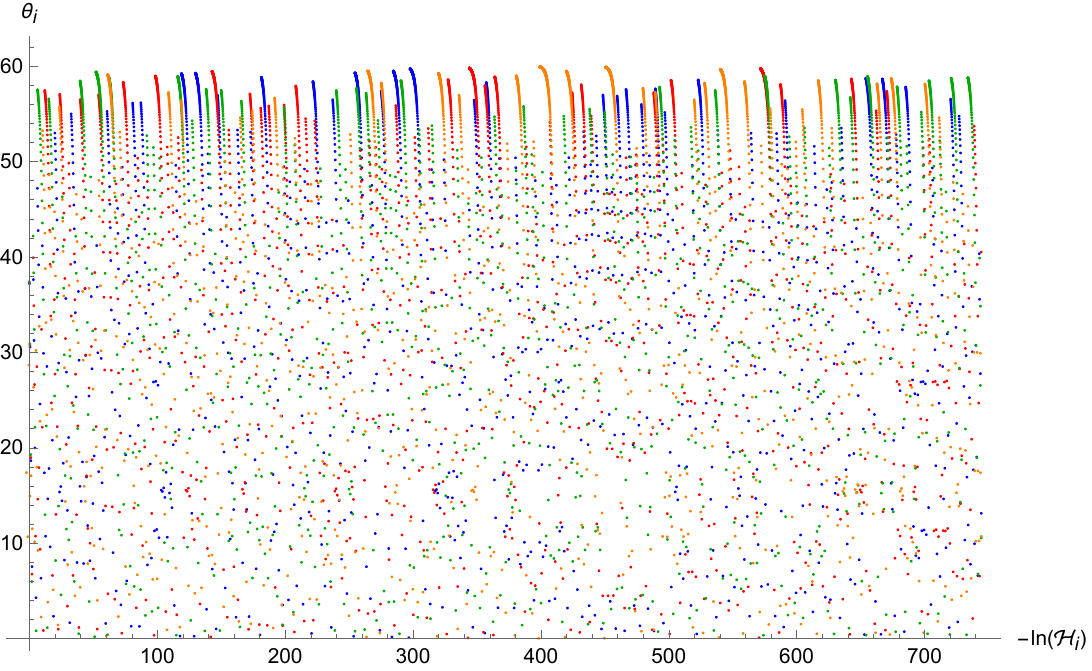}
    \caption{Trajectories %MDPI: Please confirm if explanation for color should be added. Answer: each color corresponds to a different set of initial conditions; we believe this caption makes it sufficiently clear, but let us know if we should expand on this (this is true for all figures with more than one trajectory below).
 of the point universe in the $(\mathcal{H}_i,\theta_i)$ phase space for four different sets of initial conditions in the standard case. It is evident how the point universe explores the entire available phase space showing an ergodic and hence chaotic~behavior.}
    \label{chaos}
\end{figure}

This %MDPI: We added an indentation, please check.
 result, together with the fact that the system has the symmetry of an equilateral triangle, proves that a reflection against one of the three walls is always possible. In~particular, the~number of reflections is infinite while the point universe goes towards the initial singularity. Hence, in~this picture, the only effect of the potential walls is to randomly change the direction of the free-particle motion between two subsequent reflections, thus generating chaotic dynamics (see Figure~\ref{chaos}).

\unskip
\section{Deformed Commutation~Relations}
\label{Algebras}
In this section, we present the Deformed Commutation Relations that will be later used to deform the anisotropy variables of the Bianchi IX model. The~first and most widespread one is the GUP representation in its original KMM formulation, after~the authors Kempf--Mangano--Mann~\cite{KMM}. It was developed to easily implement the concept of a minimal length, largely supposed and expected in QG theories. It consists of a deformation of the standard Heisenberg Commutation Relations, which implies a correction to the corresponding uncertainty principle:
\begin{equation}
    \comm{\hat{q}}{\hat{p}}=i\,(1+\mu^2\hat{p}^2),
\end{equation}
\begin{equation}
    2\Delta q\Delta p=1+\mu^2\Delta p^2+\mu^2\ev{\hat{p}}^2.
\end{equation}

This %MDPI: We added an indentation, please check.
 is the same uncertainty relation that appears in perturbative STs~\cite{ST1,ST2}. When considering $\ev{\hat{p}}=0$, this implies an absolute minimal uncertainty on position $\Delta q_0=\mu$. This is interpreted as a reminder that it is not possible to probe lengths smaller than the string length. However, as~will also be the case for other forms, once the deformation is implemented, the~parameter $\mu$ loses any connection to the parent theory and will just play the role of a deformation parameter; this has been explained better in~\cite{IJGMMP} in the cosmological context. Indeed, the parameter $\mu$ controls when the corrections are relevant, in~that for $\mu\to0$ (or equivalently for $\ev{\hat{p}}\ll1/\mu$), we recover the standard features. Since the KMM Algebra has already been explored in depth in~\cite{SebyMixmaster}, we will focus on other~forms.

We generalize the KMM formulation to other functions of the momentum:
\begin{equation}
    \comm{\hat{q}}{\hat{p}}=i\,f(\hat{p}).
\end{equation}

Of %MDPI: We added an indentation, please check.
 course, as~mentioned above, it is possible to use even more exotic forms, but~we will restrict ourselves to these as they usually give the kinds of cut-off expected from QG theories~\cite{Hossenfelder}.

First of all, it is important to highlight that these kinds of DCRs are better analyzed in the momentum polarization, meaning that wavefunctions will depend on the momentum as $\psi=\psi(p)$. Then, there are two main operatorial representations: one inspired by the original KMM formulation where the position operator is modified, and~one where instead the momentum operator is modified: %MDPI: Please confirm if all the sub equations should be revised. e.g., change 13a to 13 and 13b to 14. Or confirm keep the current numbering. Answer: no, this is fine.
\begin{subequations}
\begin{align}
    (1)&\quad\hat{q}\,\psi(p)=i\,f(p)\,\dv{\psi}{p},\qquad\hat{p}\,\psi(p)=p\,\psi(p),\label{modQrepresentation}\\
   (2)&\quad\hat{q}\,\psi(p)=i\,\dv{\psi}{p},\qquad\hat{p}\,\psi(p)=g(p)\,\psi(p),\label{modPrepresentation}
\end{align}
\end{subequations}
where the function $g$ is related to $f$ through $g^{-1}=\int dp/f(p)$. The~equivalence between the two representations is still unclear. They seem to yield compatible results in cosmology when applied to one-dimensional (sub)systems~\cite{IJGMMP}; however, the~equivalence seems to fail in higher spatial dimensions, since the second representation is not well defined unless some space--time symmetries are broken~\cite{Maggiore2021,SebyGUPBianchi,BarcaGielen}.

Regarding higher spatial dimensions, it is natural to generalize the DCRs as
\begin{equation}
    \comm{\hat{q}_i}{\hat{p}_j}=i\,\delta_{ij}\,f(\hat{p}_\text{tot}),\qquad p_\text{tot}^2=\sum_ip_i^2.
\end{equation}\vspace{-12pt}

In this %MDPI: We added an indentation, please check.
 way, most space--time symmetries are maintained, whereas if the function $f$ depended only on the component along that direction, the~symmetries would have been broken. This has some interesting consequences. First, as~mentioned earlier, it is impossible to define the function $g$ as in \eqref{modPrepresentation} in a consistent way, and~therefore, only the first representation \eqref{modQrepresentation} is available (the position operators $\hat{q}_i$ become derivatives with respect to the corresponding component $p_i$). Second, in~higher space dimensions, the Jacobi identities must be satisfied, and~they naturally implement a Non-Commutativity between different space directions~\cite{Maggiore2021,SebyGUPBianchi} (the components of the momentum will still commute with each other); in 3D, the commutator can be written as
\begin{equation}
    \comm{\hat{q}_i}{\hat{q}_j}=i\,A(\hat{p}_\text{tot})\,\hat{J}_{ij},
\end{equation}
where the function $A$ depends on $f$, and~$J$ is a deformed angular momentum:
\begin{equation}
    A(p)=\frac{f(p)}{p}\,\dv{f}{p},\qquad J_{ij}=\frac{q_i\,p_j-q_j\,p_i}{f(p_\text{tot})}.
\end{equation}

This %MDPI: We added an indentation, please check.
 clearly changes the dynamics and gives a possible quantum motivation for Non-Commutative effects. The~commutative case is recovered when $f=1$ and $A=0$.

In our study, we will be dealing mainly with the classical limit of these DCRs, so we need not worry about operatorial representations at this stage. Indeed, one of the most powerful features of these modified algebras is their straightforward classical limit: the DCRs just become Deformed Poisson Brackets, reading as
\begin{equation}
    \pb{q_i}{p_j}=\delta_{ij}\,f(p_\text{tot}),\qquad\pb{q_i}{q_j}=A(p_\text{tot})\,J_{ij}.
\end{equation}

This %MDPI: We added an indentation, please check.
 makes QG effects easily implementable on any kind of Hamiltonian system, and~the effective equations of motion obtain some correction factors:
\begin{equation}
    \dot{q_i}=\pb{q_i}{\ham}=\pdv{\ham}{p_i}\,f(p_\text{tot})+\sum_{j\neq i}\pdv{\ham}{q_j}\,A(p_\text{tot})\,J_{ij},\qquad\dot{p}=\pb{p}{\ham}=-\pdv{\ham}{q}\,f(p).
\end{equation}

Let %MDPI: We added an indentation, please check. Answer: this indentation can stay.
 us now present the three specific forms of DCR that we will use in our study\endnote{Note that in~\cite{IJGMMP}, the algebras had slightly different names.}.

The first one is an extension of the KMM GUP; it is called Brane Algebra because it reproduces the same modified Friedmann equation of Brane Cosmology~\cite{Battisti,IJGMMP}. It takes the form
\begin{equation}
\label{Brane}
    f_\text{Brane}(p)=\sqrt{1+\mu^2p^2\,}\,.
\end{equation}

Except %MDPI: We added an indentation, please check.
 for a factor of $2$, the~previous KMM function is basically the two leading terms of a Taylor expansion of this one, which is why it is considered an extension. However, interestingly enough, this algebra does not imply an absolute minimal uncertainty on position. The~issue is that the square root is a non-local function: when computing the uncertainty relations, we need to perform a Taylor expansion, make the operator act in series form, and~then re-sum, but~the resummation has a finite convergence radius that does not allow for the same conclusions as the KMM case. The~minimal uncertainty is then recovered by imposing an artificial momentum cut-off by hand~\cite{SebyGUPmomentumcutoff}.

The second form is called Loop Algebra because it is inspired by LQC. It takes the~form
\begin{equation}
\label{Loop}
    f_\text{Loop}(p)=\sqrt{1-\mu^2p^2\,}\,.
\end{equation}

In one %MDPI: We added an indentation, please check.
 space dimension, the~corresponding momentum operator in the representation \eqref{modPrepresentation} would be
\begin{equation}
    \hat{p}\,\psi(p)=\frac{\sin(\mu\,p)}{\mu}\,\psi(p),
\end{equation}
which is the same operator as LQC~\cite{BojowaldOriginalLQC,LQC2011Review}. Indeed, it is able to reproduce exactly the same dynamics as effective LQC: it clearly introduces a momentum cut-off as $\ev{\hat{p}^2}\leq1/\mu^2$ (otherwise, the commutator would not be purely imaginary anymore and unitarity would be lost). When implemented on the volume (i.e., the cubed scale factor) and its conjugate momentum, which is directly related to energy density and the Hubble parameter through the Hamiltonian constraint, it is able to introduce a cut-off and replace cosmological singularities with a Bounce~\cite{IJGMMP}.

The third and final form is called the Loop Uncertainty Principle (LUP), and~can be seen as either the counterpart of KMM with a minus sign (indeed, sometimes in the literature, the two are considered together~\cite{Battisti}), or~as the leading terms of the Taylor-expanded Loop Algebra. It reads as
\begin{equation}
    \label{LUP}
    f_\text{LUP}(p)=1-\mu^2p^2.
\end{equation}

In the %MDPI: We added an indentation, please check.
 cosmological context, this algebra has also been able to remove singularities, but~instead of a loop-like Bounce, it introduces an asymptotically Einstein-static phase, reproducing what is known as an emergent universe~\cite{IJGMMP,BarcaEU}. %EE: please check that the intended meaning has been retained. Answer: if possible, it would be better to keep the capital letter for Emergent Universe, since it is the proper name of a model.

In our study, we will implement these three DCRs on the anisotropy variables $\beta_\pm$, which will form a (deformed) two-dimensional space and will therefore not commute between each other anymore:
\begin{subequations}
\begin{equation}
    \pb{\beta_\pm}{p_\pm}=\delta_\pm\,f(p_\text{tot}),\qquad p_\text{tot}^2=p_+^2+p_-^2,
\end{equation}
\begin{equation}
    \pb{\beta_+}{\beta_-}=A(p_\text{tot})\,\frac{\beta_+p_--\beta_-p_+}{f(p_\text{tot})}.
\end{equation}
\end{subequations}

As in %MDPI: We added an indentation, please check.
 the classical picture, the~isotropic variable $\alpha$ will play the role of time, and~therefore will be left~undeformed.
\section{Bianchi Models with Deformed Poisson~Algebras}
\label{DeformedBianchi}
The aim of this section is to analyze how the Mixmaster dynamics, i.e.,~the picture of the point particle in a triangular box that falls into the initial singularity, are modified when we consider DCRs in the $(\beta_+,\beta_-,p_+,p_-)$ phase space. As~already mentioned, the~motion of the point universe for $\alpha\rightarrow-\infty$ is that of a free particle, except~when it is reflected against a wall. Hence, instead of considering the full Bianchi IX potential, it is possible to consider one wall at a time and use the triangular symmetry to rotate the system, in~order to iterate the dynamics. This approximation corresponds to considering the Bianchi II model, whose  Hamiltonian is
\begin{equation}
\mathcal{H}^{II}=\sqrt{p_+^2+p_-^2+3(4\pi^4)e^{4(\alpha-2\beta_+)}}\,,
\end{equation}
in which we choose only one of the three terms of the full potential \eqref{V}. This picture describes exactly one single reflection against a potential~wall.

Now, we want to study the properties of the dynamics in the framework of DCRs
\begin{subequations}
\begin{equation}
\pb{p_+}{p_-}=0\,,
\end{equation}
\begin{equation}
\pb{\beta_+}{\beta_-}=\displaystyle(\beta_+\,p_--\beta_-\,p_+)\frac{f'(p_{\text{tot}})}{p_{\text{tot}}}\,,
\end{equation}
\begin{equation}
\pb{\beta_\pm}{p_\pm}=\delta_\pm f(p_{\text{tot}})\,.
\end{equation}
\end{subequations}

Our %MDPI: We added an indentation, please check.
 interest is understanding if chaos survives in this context. Therefore, we have to construct the modified integrals of motion and then find the modified reflection law that governs how the directions of the point universe before and after a reflection are mapped. In~particular, from~the modified Hamilton equations
\begin{equation}
    \begin{cases}
\dot{p}_{+}=\displaystyle3(4\pi^4)f(p)\frac{e^{4(\alpha-2\beta_+)}}{\mathcal{H}^{II}}\,,\\
\dot{p}_{-}=0\,,\\
\dot{\beta}_{+}=\displaystyle\frac{p_{+}}{\mathcal{H}^{II}}f(p)\,,\\
\dot{\beta}_{-}=\displaystyle\frac{p_{-}}{\mathcal{H}^{II}}f(p)+3(4\pi)^4\frac{(\beta_+\,p_--\beta_-\,p_+)f'(p_{\text{tot}})}{p_{\text{tot}}}\frac{e^{4(\alpha-2\beta_+)}}{\mathcal{H}^{II}}\,,\\
\dot{\mathcal{H}}^{II}=\displaystyle\frac{6(4\pi)^4e^{4(\alpha-2\beta_+)}}{\mathcal{H}^{II}}\,,
    \end{cases}
\end{equation}
it is easy to see that the constants of motion are
\begin{equation}
    \label{const}
        p_-\,,\quad\displaystyle\mathcal{H}-\frac{1}{2}\int_{\bar{p}_+}^{p_+}\frac{1}{f(p_\text{tot})}dp_+'\,.
    \end{equation}
    
In order %MDPI: We added an indentation, please check. Answer: this indentation can stay.
 to find the modified point-universe velocity, we go back to the Bianchi I approximation, in~which the Hamilton equations are
\begin{equation}
    \begin{cases}
\dot{\mathcal{H}}^{I}=\dot{p}_{+}=\dot{p}_{-}=0\\
\dot{\beta}_{\pm}=\displaystyle\frac{p_{\pm}}{\mathcal{H}^{I}}f(p_{\text{tot}})\\
    \end{cases}
\end{equation}
and from which we obtain $\dot{\beta}=f(p_\text{tot})$. Therefore, in principle, it is not granted that the point particle is always faster than the potential walls. In~particular, it can be demonstrated that the wall velocity remains unchanged. Indeed, the~condition for the wall to be relevant is $\mathcal{H}^{I}/e^{4(\alpha-2\beta_-)}\sim1$ and it involves only the definition of the Hamiltonian, which is unaffected by DCRs. Hence, we find again $\dot{\beta}_{\text{wall}}=\frac{1}{2}$.

Now, we introduce a convenient parametrization to study the maximum incident angle. In~particular, we define
\begin{subequations}
\label{par1}
\begin{equation}
(p_-)_{i,f}=\mathcal{H}^{I}_{i,f}\sin\theta_{i,f}
\end{equation}
\begin{equation}
(p_+)_i=-\mathcal{H}^{I}_i\cos\theta_i\,,\quad (p_+)_f=\mathcal{H}^{I}_f\cos\theta_f\,,
\end{equation}
\end{subequations}
from which we also obtain 
\begin{subequations}
\label{par2}
\begin{equation}
(\dot{\beta}_-)_{i,f}=f(p_{\text{tot}})\sin\theta_{i,f}\,,
\end{equation}
\begin{equation}
(\dot{\beta}_+)_i=-f(p_{\text{tot}})\cos\theta_i\,,\quad (\dot{\beta}_+)_f=f(p_{\text{tot}})\cos\theta_f\,.
\end{equation}
\end{subequations}

Therefore, %MDPI: We added an indentation, please check.
 from~the condition $(\dot{\beta}_+)_i>\dot{\beta}_{\text{wall}}$, we find
\begin{equation}
\label{cond}
\theta_i<\displaystyle\arccos\bigg(\frac{1}{2f(p_{\text{tot}})}\bigg)\,,
\end{equation}
that is compatible with $\theta_{max}=\frac{\pi}{3}$ in the standard case $f(p_{\text{tot}})=1$. 

At this point, the~strategy to investigate the dynamics is as follows. By~using the two constants of motion \eqref{const}, we obtain the deformed reflection law for a generic $f(p_{\text{tot}})$:
\begin{equation}
\label{BKL}
\begin{cases}
    \mathcal{H}_i\sin{\theta_i}=\mathcal{H}_f\sin{\theta_f}\,,\\
    \displaystyle\mathcal{H}^i-\frac{1}{2}\int_{\bar{p}_+}^{p_+^i}\frac{1}{f(p_+',p_-)}\,dp_+'=\mathcal{H}^f-\frac{1}{2}\int_{\bar{p}_+}^{p_+^f}\frac{1}{f(p_+',p_-)}\,dp_+'\,,
\end{cases}
\end{equation}
in which we have used the parametrization \eqref{par1} in defining $(p_-)^{i,f}$. Once the initial conditions $(\theta_i,\mathcal{H}_i)$ have been fixed, the~map gives the final values $(\theta_f,\mathcal{H}_f)$ after the reflection. Then, in~order to completely solve the dynamics, the~procedure is iterated using the relation between the final angle $\theta_f$ and the new initial angle $\theta_i'$
\begin{equation}
\label{thetai}
\begin{cases}
    \theta_i'=\frac{\pi}{3}-\theta_f\quad\quad\text{if}\quad\theta_f\leq\frac{\pi}{3}\,,\\
    \theta_i'=-\frac{\pi}{3}+\theta_f\quad\,\text{if}\quad\theta_f>\frac{\pi}{3}\,,
\end{cases}
\end{equation}
that can be easily found by using the triangular symmetry of the system. Differently from the standard picture (see Section~\ref{BIX}), the~condition \eqref{cond} is not automatically matched after each reflection  by the iterated initial angle $\theta_i'$, so it is not given that chaos is still present in the deformed~picture.  

In the next sections, we will proceed separately for the chosen DCRs, as~mentioned in Section~\ref{Algebras}, and~finally we will discuss the~results.

\subsection{Brane~Algebra}
Here, we implement the Brane Algebra $f_{\text{Brane}}(p)$ as presented in \eqref{Brane}, thus obtaining the following modified symplectic structure for the two anisotropic degrees of freedom $(\beta_+,\beta_-)$
\begin{subequations}
\begin{equation}
\pb{p_+}{p_-}=0\,,
\end{equation}
\begin{equation}
\pb{\beta_+}{\beta_-}=\displaystyle\frac{\mu^2\,(\beta_+p_--\beta_-p_+)}{\sqrt{1+\mu^2(p_+^2+p_-^2)\,}\,} \,,
\end{equation}
\begin{equation}
\pb{\beta_\pm}{p_\pm}=\delta_\pm\sqrt{1+\mu^2(p_+^2+p_-^2)\,}\,.
\end{equation}
\end{subequations}

Hence, %MDPI: We added an indentation, please check.
 the~Hamilton equations with these modified Poisson brackets in the Bianchi I approximation are
\begin{equation}
\begin{cases}
\dot{p}_{\pm}=0\,,\\
\dot{\beta}_{\pm}=\displaystyle\frac{p_{\pm}}{\sqrt{p_+^2+p_-^2}}\sqrt{1+\mu^2(p_+^2+p_-^2)}\,,
\end{cases}
\end{equation}
from which we obtain the modified point-particle velocity, i.e.,
\begin{equation}
\dot{\beta}=\sqrt{\dot{\beta}_+^2+\dot{\beta}_-^2}=\sqrt{1+\mu^2(p_+^2+p_-^2)}>1\,.
\end{equation}

\begin{figure}[H]
   % \centering 
    \includegraphics[width=0.79\linewidth]{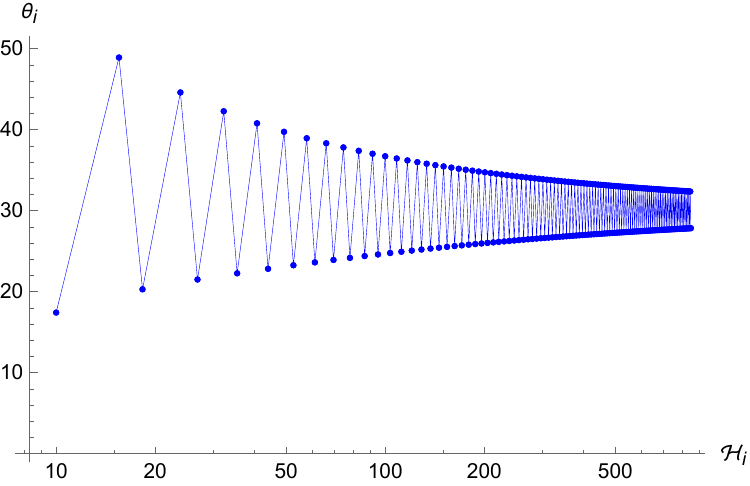}
    \caption{Trajectory of the point universe in the $(\mathcal{H}_i,\theta_i)$ phase space for the Brane Algebra. The~energy $\mathcal{H}_i$ keeps increasing (differently from the standard case where it always decreases), whereas the angle $\theta_i$ oscillates indefinitely between two converging~values.}
    \label{GUPb1}
\end{figure}

\begin{figure}[H]
   % \centering
\includegraphics[width=0.89\linewidth]{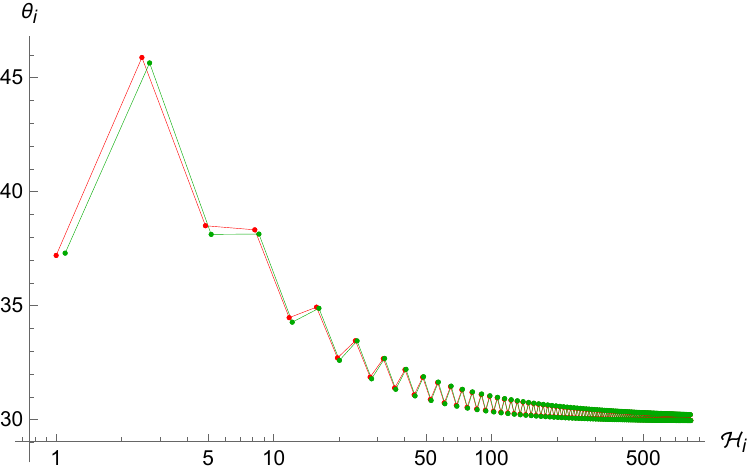}
    \caption{Trajectories %MDPI: Please confirm if explanation for color should be added.
 of the point universe in the $(\mathcal{H}_i,\theta_i)$ phase space for two different (similar) sets of initial conditions in the Brane Algebra. It is evident that the two trajectories are not sensitive to the initial~conditions.}
    \label{GUPb2}
\end{figure}

\begin{figure}[H]  
\begin{adjustwidth}{-\extralength}{0cm}
  \centering
\includegraphics[width=0.42\paperwidth]{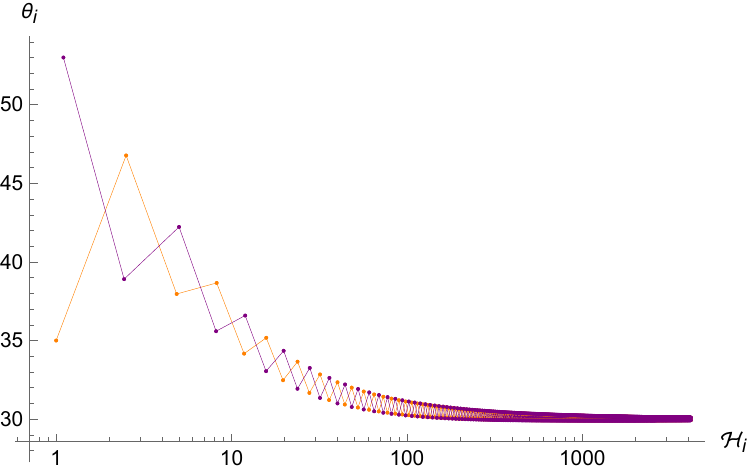}\,\,\includegraphics[width=0.42\paperwidth]{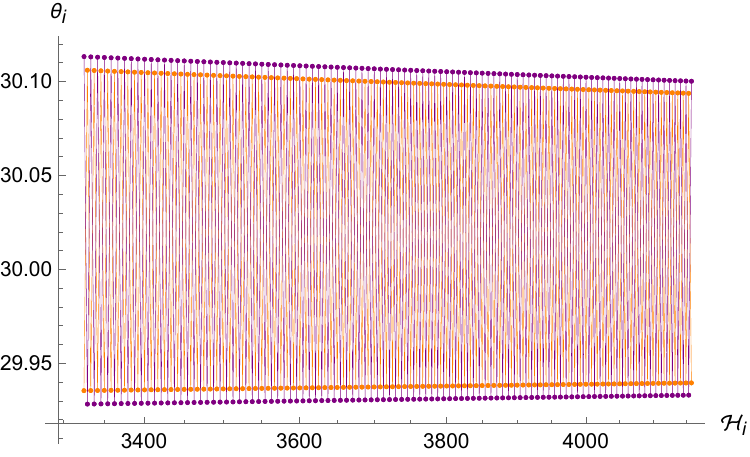}
\end{adjustwidth}
    \caption{Left: trajectories %MDPI: Please confirm if explanation for color should be added.
 of the point universe in the $(\mathcal{H}_i,\theta_i)$ phase space for two (sufficiently) different sets of initial conditions in the Brane Algebra. After~thousands of iterations, both converge to the same angle of $\pi/6$. Right: same figure, zoomed to the last few hundred~iterations.}
    \label{GUPb3-20000}
\end{figure}

As we %MDPI: We added an indentation, please check.
 previously mentioned, the~wall velocity is still equal to $\frac{1}{2}$, whereas in this particular algebra, the point particle becomes faster. In~principle, this property could suggest the idea that chaos is enhanced in the Brane Algebra framework. To~verify this hint, we proceed with the analysis of the BKL map \eqref{BKL}, which in this Algebra becomes
\begin{equation}
\label{BKLGUP}
\begin{cases}
\ham_i\,\sin\theta_i=\ham_f\,\sin\theta_f\,,\\
\displaystyle\ham_i+\frac{1}{2\mu}\text{arctanh}\left(\frac{\mu\,\ham_i\cos\theta_i}{\sqrt{1+\mu^2\ham_i^2\,}\,}\right)=\ham_f-\frac{1}{2\mu}\text{arctanh}\left(\frac{\mu\,\ham_f\cos\theta_f}{\sqrt{1+\mu^2\ham_f^2\,}\,}\right)\,,
\end{cases}
\end{equation}
in which we have used the parametrization \eqref{par1} for the definition of $(p_\pm)^{i,f}$. Differently from the standard case \eqref{reflection}, this system cannot be solved analytically. Therefore, we need to implement numerical methods to solve the~dynamics.    In particular, %MDPI: Please confirm if it  should be split from here (new paragraph). Answer: no, we prefer to have all of this as one paragraph.
 Figures~\ref{GUPb1} and \ref{GUPb2} show the trajectories in the $(\mathcal{H}_i,\theta_i)$ phase space for different initial conditions. As~one can see from \eqref{cond} and \eqref{thetai}, the~condition for a reflection is always met when $f(p_\text{tot})=f_\text{Brane}(p_\text{tot})$ and the point particle experiences an infinite series of reflections while falling into the initial singularity. However, the~motion loses its chaotic feature and sets into an oscillatory orbit. Indeed, as~one can see from \mbox{Figures~\ref{GUPb2} and \ref{GUPb3-20000}}, the~trajectories tend to become very close to each other for both very similar and sufficiently different initial conditions. In~particular, they seem to oscillate ever closer to the value $\pi/6$ which, thanks to the combination of the first equation of \eqref{thetai} and the structure of the map \eqref{BKLGUP} for large values of $\ham$, becomes an attractor value.

\unskip
\unskip

%\finishcolumns
\unskip
%\startcolumns
%\switchcolumn

\subsection{Loop~Algebra}
\label{LoopAlgebra}
Now, we study the motion of the point universe in the Algebra inspired by LQC $f_{\text{Loop}}$ (see \eqref{Loop}), for~which the modified algebra for the anisotropies reads as
\begin{subequations}
\begin{equation}
\pb{p_+}{p_-}=0\,,
\end{equation}
\begin{equation}
\pb{\beta_+}{\beta_-}=\displaystyle-\frac{\mu^2\,(\beta_+p_--\beta_-p_+)}{\sqrt{1-\mu^2(p_+^2+p_-^2)\,}\,} \,,
\end{equation}
\begin{equation}
\pb{\beta_\pm}{p_\pm}=\delta_\pm\sqrt{1-\mu^2(p_+^2+p_-^2)\,}\,,
\end{equation}
\end{subequations}
and the Hamilton equations in the Bianchi I approximation are
\begin{equation}
\begin{cases}
\dot{p}_{\pm}=0\,,\\
\dot{\beta}_{\pm}=\displaystyle\frac{p_{\pm}}{\sqrt{p_+^2+p_-^2}}\sqrt{1-\mu^2(p_+^2+p_-^2)}\,.
\end{cases}
\end{equation}

While $\dot{\beta}_{\text{wall}}$ remains unchanged, for~the point-universe velocity, we find
\begin{equation}
\dot{\beta}=\sqrt{\dot{\beta}_+^2+\dot{\beta}_-^2}=\sqrt{1-\mu^2(p_+^2+p_-^2)}<1
\end{equation}

Hence, %MDPI: We added an indentation, please check.
 the~condition \eqref{cond} is not always met. In~particular, when that happens, the~iteration map stops, the~reflections end, and the point particle keeps moving indefinitely with a free motion towards the initial singularity. For~this reason, we expect chaos to be suppressed in this~scenario. 

As in the previous case, we write down the modified BKL map for the considered Algebra, i.e.,
\begin{equation}
\label{BKLLOOP}
\begin{cases}
\ham_i\,\sin\theta_i=\ham_f\,\sin\theta_f\,,\\
\displaystyle\ham_i+\frac{1}{2\mu}\arctan(\frac{\mu\,\ham_i\cos\theta_i}{\sqrt{1-\mu^2\ham_i^2\,}\,})=\ham_f-\frac{1}{2\mu}\arctan(\frac{\mu\,\ham_f\cos\theta_f}{\sqrt{1-\mu^2\ham_f^2\,}\,})\,,
\end{cases}
\end{equation}
and we study the properties of the motion by numerically solving the iterative system, once given initial conditions $(\mathcal{H}_i,\theta_i)$ and checking at every step that condition \eqref{cond} is satisfied. In~the below figures, trajectories for different initial  conditions are~represented. 

\begin{figure}[H]
 %   \centering
    \includegraphics[width=0.9\linewidth]{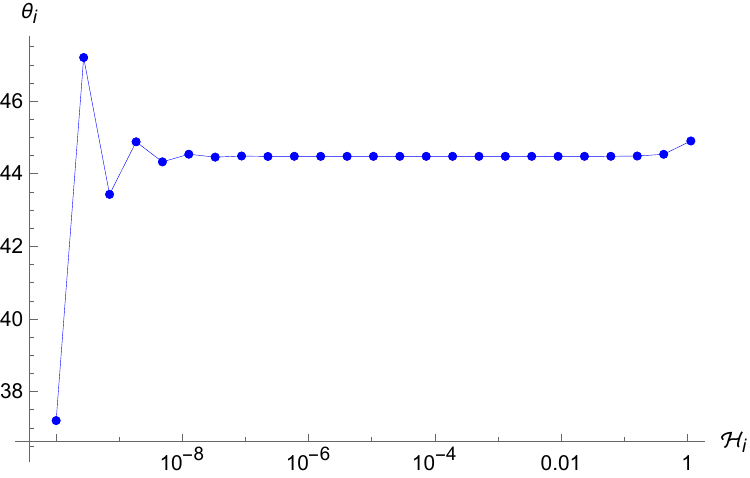}
    \caption{Trajectory of the point universe in the $(\mathcal{H}_i,\theta_i)$ phase space for the Loop Algebra. The~energy $\mathcal{H}_i$ keeps decreasing whereas the angle $\theta_i$ oscillates between two close values, before~stopping after 25~iterations.}
    \label{PUPe1}
\end{figure}

\begin{figure}[H]
    %\centering
     \includegraphics[width=0.9\linewidth]{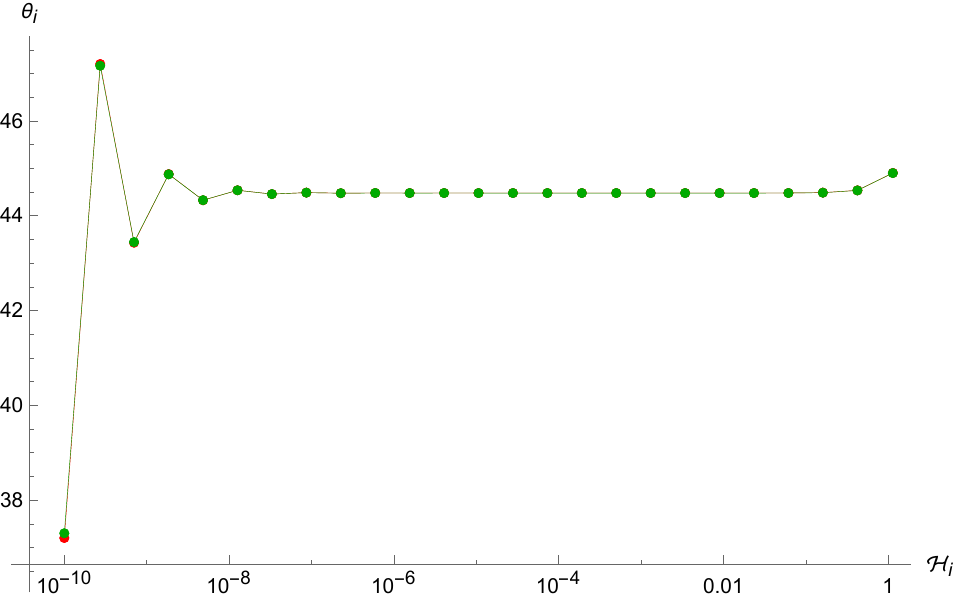}
    \caption{Trajectories %MDPI: Please confirm if explanation for color should be added.
 of the point universe in the $(\mathcal{H}_i,\theta_i)$ phase space for two different (similar) sets of initial conditions in the Loop Algebra. It is evident that the two trajectories are not sensitive to the initial~conditions.}
    \label{PUPe2}
\end{figure}

\begin{figure}[H]
 %   \centering
    \includegraphics[width=0.9\linewidth]{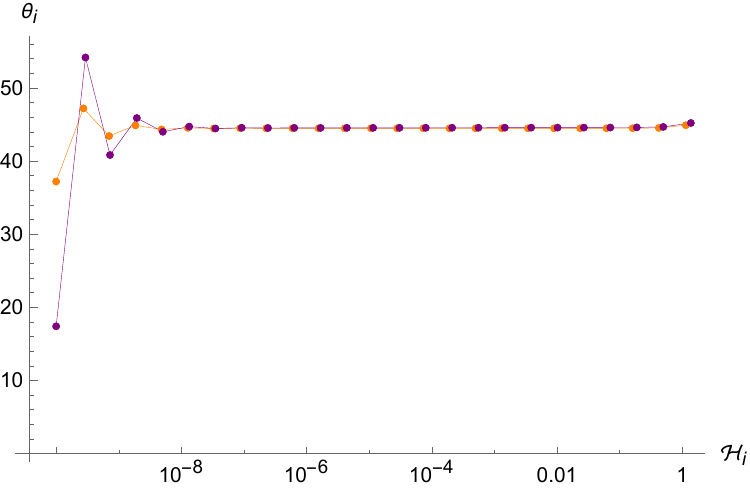}
    \caption{Trajectories %MDPI: Please confirm if explanation for color should be added.
 of the point universe in the $(\mathcal{H}_i,\theta_i)$ phase space for two different sets of initial conditions in the Loop Algebra. The~two trajectories seem to converge even if they started far from each~other.}
    \label{PUPe3}
\end{figure}

In particular, it is even more evident in this case the ergodicity is completely lost with respect to the Brane Algebra scenario (see Figures~\ref{PUPe1}--\ref{PUPe3}). As~anticipated above, the~system tends again to oscillate between two very close values of $\theta_i$, even with very different choices of initial conditions, before~stopping after a finite number of iterations when the term $\mu^2\ham^2$ is of order unity, which makes the point-universe velocity very small. Furthermore, we see again that there is an attractor value, this time at $\pi/4$. This is probably due to the fact that in order to observe enough reflections, we must give a very small initial condition for $\ham_i$; therefore, we can expand the arctangents around zero and simplify the map \eqref{BKLLOOP}, which for $\theta_i\approx\pi/4$ yields $\theta_f\approx\pi/12$, and thanks to \eqref{thetai}, the next $\theta_i'$ will again be around $\pi/4$.

\unskip

\unskip

\subsection{LUP~Algebra}
Finally, we analyze the properties of the point-universe dynamics in the LUP Algebra, where $f_{\text{LUP}}$ (see \eqref{LUP}) contains only the leading terms of the expansion of $f_{\text{Loop}}$, as~discussed in Section~\ref{Algebras}. In~this framework, the~modified algebra for the phase space of the anisotropies~is
\begin{subequations}
\begin{equation}
\pb{p_+}{p_-}=0\,,
\end{equation}
\begin{equation}
\pb{\beta_+}{\beta_-}=\displaystyle-2\mu^2\,(\beta_+p_--\beta_-p_+)\,,
\end{equation}
\begin{equation}
\pb{\beta_\pm}{p_\pm}=\delta_\pm\Bigl(1-\mu^2(p_+^2+p_-^2)\,\Bigr)\,,
\end{equation}
\end{subequations}
and the corresponding Hamilton equations in the Bianchi I approximation are
\begin{equation}
\begin{cases}
\dot{p}_{\pm}=0\,,\\
\dot{\beta}_{\pm}=\displaystyle\frac{p_{\pm}}{\sqrt{p_+^2+p_-^2}}\Bigl(1-\mu^2(p_+^2+p_-^2)\Bigr)\,.
\end{cases}
\end{equation}

Similarly %MDPI: We added an indentation, please check.
 to what happens with the Loop Algebra, the~point universe is slower, as~the following relation shows:
\begin{equation}
\dot{\beta}=\sqrt{\dot{\beta}_+^2+\dot{\beta}_-^2}=\Bigl(1-\mu^2(p_+^2+p_-^2)\Bigr)<1\,. 
\end{equation}

Hence, %MDPI: We added an indentation, please check.
 with~similar motivations to those explained above, we expect the suppression of chaos once the condition \eqref{cond} is not satisfied anymore by the constants of motion of the point particle. Then, the~reflections stop and the point universe proceeds towards the initial singularity with no more reflections on the potential~walls. 

\begin{figure}[H]
   % \centering
    \includegraphics[width=0.9\linewidth]{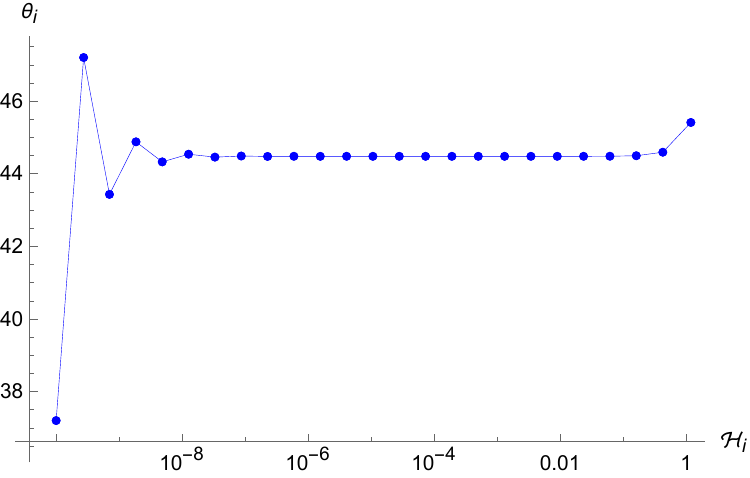}
    \caption{Trajectory of the point universe in the $(\mathcal{H}_i,\theta_i)$ phase space for the LUP Algebra. The~energy $\mathcal{H}_i$ keeps decreasing whereas the angle $\theta_i$ oscillates between two close values, before~stopping after 25~iterations.}
    \label{PUPt1}
\end{figure}

\begin{figure}[H]
 %   \centering
  \includegraphics[width=0.9\linewidth]{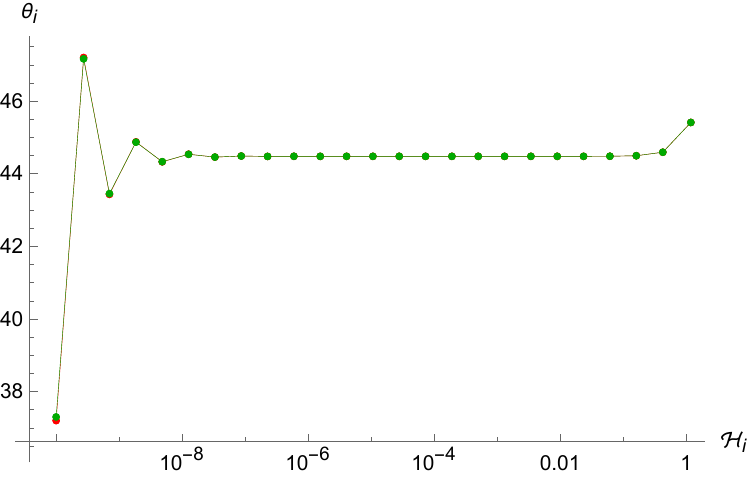}
    \caption{Trajectories %MDPI: Please confirm if explanation for color should be added.
 of the point universe in the $(\mathcal{H}_i,\theta_i)$ phase space for two different (similar) sets of initial conditions in the LUP Algebra. It is evident that the two trajectories are not sensitive to the initial~conditions.}
    \label{PUPt2}
\end{figure}

\begin{figure}[H]
   % \centering
    \includegraphics[width=0.9\linewidth]{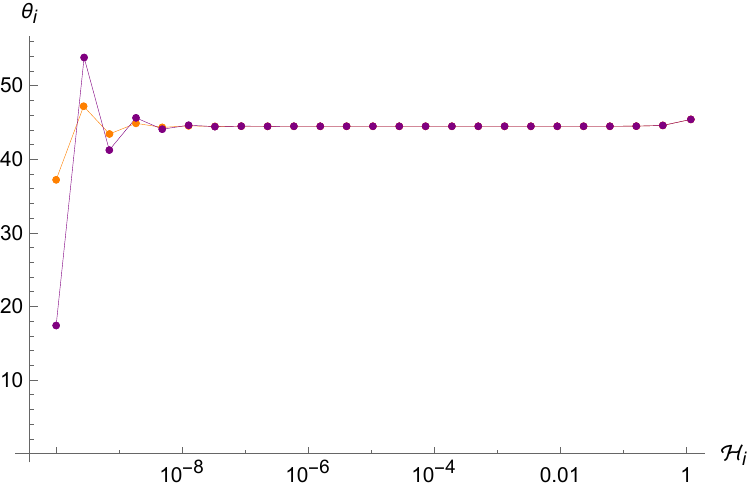}
    \caption{Trajectories %MDPI: Please confirm if explanation for color should be added.
 of the point universe in the $(\mathcal{H}_i,\theta_i)$ phase space for two different sets of initial conditions in the LUP Algebra. The~two trajectories seem to converge even if they started far from each~other.}
    \label{PUPt3}
\end{figure}

In order to show the trajectories of the point universe in this scenario, we numerically study the modified BKL map
\begin{equation}
\label{BKLLUP}
\begin{cases}
\ham_i\,\sin\theta_i=\ham_f\,\sin\theta_f\,,\\
\displaystyle\ham_i+\frac{\text{arctanh}\left(\frac{\mu\,\ham_i\cos\theta_i}{\sqrt{1-\mu^2\ham_i^2\sin^2\theta_i\,}\,}\right)}{2\mu\sqrt{1-\mu^2\ham_i^2\sin^2\theta_i\,}\,}=\ham_f-\frac{\text{arctanh}\left(\frac{\mu\,\ham_f\cos\theta_f}{\sqrt{1-\mu^2\ham_i^2\sin^2\theta_i\,}\,}\right)}{2\mu\sqrt{1-\mu^2\ham_i^2\sin^2\theta_i\,}\,}\,.
\end{cases}
\end{equation}
from which we obtain the following trajectories for different sets of initial conditions (\mbox{Figures \ref{PUPt1}--\ref{PUPt3}}). 

Also in this case, the~point universe follows oscillatory trajectories before settling into the final uniform rectilinear motion until the singularity is reached. In~particular, the~ergodic property of the standard motion (see Figure~\ref{chaos}) is lost, since trajectories related to similar initial conditions remain close to each other. Analogously to the loop case of Section~\ref{LoopAlgebra}, here, we also find an attractor at $\pi/4$. This is to be expected because the hyperbolic arctangents of the map \eqref{BKLLUP} have the same leading-order behavior for very small arguments of the loop map \eqref{BKLLOOP}, and therefore, the chain $\theta_i\approx\pi/4$, $\theta_f\approx\pi/12$, $\theta_i'\approx\pi/4$ is also reproduced~here.

\unskip
\unskip
\section{Conclusions}
\label{Concl}
In this study, we analyzed the Bianchi IX model in the Misner variables by implementing DCRs for the two anisotropic variables $(\beta_+,\beta_-)$. As~shown in~\cite{Misner1,Misner2,BKL}, the~Bianchi IX universe behaves as a chaotic dynamical system when it approaches the initial singularity, and~indeed, the aim of this study was to investigate if chaos survives when a deformed scenario is considered. As~demonstrated in~\cite{GiovannettiMixmaster}, when the initial singularity is still present (as is the case with the isotropic Misner variable, except~when pure quantum approaches are considered~\cite{Giovannetti_2022,Giovannetti_2023}), the only way to affect the chaotic properties of the dynamics is to introduce some kind of cut-off on the real gravitational degrees of freedom, i.e.,~the anisotropies. Here, we considered three different forms of DCRs inspired by STs and LQC, in~order to study how chaos behaves in an effective QG framework. This represents a first step towards the understanding of the quantum properties of the initial singularity, given that the Bianchi IX model can be considered as a building block for the generic cosmological solution~\cite{PC,BKL}.

In our analysis, we implemented the Brane, Loop and LUP Algebras, as~presented in Section~\ref{Algebras}, on~the $(\beta_+,\beta_-,p_+,p_-)$ phase space, obtaining non-trivial Poisson brackets $\{\beta_\pm,p_\pm\}$ and also Non-Commutativity between the configurational variables $(\beta_+,\beta_-)$. Obviously, these features resulted in modified Hamilton equations for the Bianchi IX model. In~particular, in~order to study the dynamics, we first used the Bianchi II approximation by exploiting the triangular symmetry of the system, and~then we implemented the Bianchi I approximation since near the initial singularity the potential becomes more and more negligible. With~these hypotheses, we found the modified reflection laws in all three cases, i.e.,~the modified BKL maps which encode the chaotic properties of the dynamics. Differently from the standard case, numerical methods were needed to solve the dynamics, and~the interesting result is that chaos does not survive in any of the three scenarios. In~particular, in~the case of Loop and LUP Algebras, the~velocity of the point universe is damped, and therefore, sooner or later it will become slower than the potential walls and there will be no more reflections. This is the effective imprint of the underlying quantum theory, i.e.,~LQC, in~which the deformation parameter $\mu$ corresponds to the non-zero step of the lattice implemented on the anisotropies $(\beta_+,\beta_-)$, and~hence introduces a maximum value for the momenta $(p_+,p_-)$~\cite{LQC2011Review,Review}. 

On the other hand, for~the Brane Algebra, the point-universe velocity is greater than in the standard case. However, this is not sufficient to infer the preservation of chaos, and~indeed the behavior of the BKL map is similar to that in~\cite{SebyMixmaster}, where the DCR contains only the leading terms of the expansion. The~parent theory of these two Algebras, i.e.,~ST, considers the string scale as the minimal measurable length; when referring to the anisotropies $(\beta_+,\beta_-)$, this minimal uncertainty implies that on a quantum level, the corners between the walls are widened and the probability of escaping from the potential well is enhanced~\cite{SebyMixmaster} (of course, this property must be properly studied in the full quantum regime). From~this perspective, it would be interesting to analyze how chaos is modified when the effects of QG theories are introduced directly at the level of the Hamiltonian, without~acting on the structure of the Poisson Algebra, similarly to~\cite{IJGMMP}.

So far, the~relevance of this analysis lies in the possibility of inferring the semiclassical properties of the Mixmaster dynamics in the context of BC and LQC using the Deformed Algebra framework, especially focusing on the fate of chaos. In~particular, it seems to be a strong classical property that can be easily suppressed by quantum gravitational~effects.

\vspace{6pt}
\authorcontributions{All authors contributed equally to all parts of the study and of the manuscript. All authors have read and agreed to the published version of the manuscript.}

\funding{This work is partially supported by the MUR PRIN Grant 2020KR4KN2 ``String Theory as a bridge between Gauge Theories and Quantum Gravity'', by~the FARE programme (GW-NEXT, CUP:~B84I20000100001), and~by the INFN TEONGRAV initiative.}

\dataavailability{No new data were created or analyzed in this study. Data sharing is not applicable to this article.}

\acknowledgments{The authors thank S. Segreto for useful insight and~discussions.}

\conflictsofinterest{The authors declare no conflicts of~interest.}%MDPI: Please confirm the statement. Answer: confirm.

\begin{adjustwidth}{-\extralength}{0cm}

\printendnotes[custom]

\reftitle{References}
%\bibliography{bibl}

\PublishersNote{}
\end{adjustwidth}

\end{document}